\documentclass{svmult}

\usepackage{multicol}    

\begin{document}

\title*{An Introduction to Lattice Chiral Fermions}
\author{Herbert Neuberger}
\institute{Department of Physics and Astronomy, Rutgers University, Piscataway, NJ08540, USA
\texttt{neuberg@physics.rutgers.edu}}

\maketitle

This write-up starts by introducing lattice chirality to people
possessing a fairly modern mathematical background, but little prior
knowledge about modern physics. I then proceed to present two
new and speculative ideas.

\section{Review}

\subsection{What are Dirac/Weyl fermions ?}

One can think about (Euclidean) Field Theory as of an attempt to
define integrals over function spaces~\cite{PandS}. The functions are of
different types and are called fields. The integrands consist of a
common exponential factor multiplied by various monomials in the
fields. The exponential factor is written as $\exp (S)$ where the
action $S$ is a functional of the fields. Further restrictions on
$S$ are: (1) locality (2) symmetries. Locality means that  $S$ can
be written as an integral over the base space (space-time) which
is the common domain of all fields and the integrand at a point
depends at most exponentially weakly on fields at other, remote,
space-time points. $S$ is required to be invariant under an all
important group of symmetries that act on the fields. In a sense,
$S$ is the simplest possible functional obeying the symmetries and
generically represents an entire class of more complicated
functionals, which are equivalently appropriate for describing the
same physics.

Dirac/Weyl fields have two main characteristics: (1) They are
Grassmann valued, which means they are anti-commuting objects and
(2) there is a form of $S$, possibly obtained by adding more
fields, where the Dirac/Weyl fields, $\psi$, enter only
quadratically. The Grassmann nature of $\psi$ implies that the
familiar concept of integration needs to be extended. The
definition of integration over Grassmann valued fields is
algebraic and for an $S$ where the $\psi$ fields enter
quadratically, as in $S=\bar\psi K \psi +....$, requires only the
propagator, $K^{-1}$, and the determinant, $\det K$. Hence, only
the linear properties of the operator $K$ come into play, and
concepts like a ``Grassmann integration measure'' are, strictly
speaking, meaningless, although they make sense for ordinary,
commuting, field integration variables.

Let us focus on a space-time that is a a 4D Euclidean flat
four torus, with coordinates $x_\mu , ~\mu=1,2,3,4$.
Introduce the quaternionic basis $\sigma_\mu$ represented
by $2\times 2$ matrices:
\begin{equation}
\sigma_1=\pmatrix{0&1\cr 1&0}~\sigma_2=\pmatrix{0&-\imath\cr
\imath&0}~\sigma_3=\pmatrix{-1&0\cr 0&1}~
\sigma_4=\pmatrix{\imath&0\cr 0&\imath}
\end{equation}
The $\psi$ fields are split into two kinds, $\bar\psi$ and $\psi$, 
each being a two component function on the torus. In the absence
of other fields the Weyl operators playing the role of the kernel
$K$ are $W=\sigma_\mu\partial_\mu$ and
$W^\dagger=-\sigma_\mu^\dagger\partial_\mu$. The Dirac operator is
made by combining the Weyl operators:
\begin{equation}
D=\pmatrix{0&W\cr -W^\dagger & 0} =
\pmatrix{0&\sigma_\mu\cr\sigma_\mu^\dagger & 0} \partial_\mu
\equiv \gamma_\mu \partial_\mu = -D^\dagger
\end{equation}
The $\sigma_\mu$ obey
\begin{equation}
\sigma_\mu^\dagger \sigma_\nu +\sigma_\nu^\dagger \sigma_\mu = 2\delta_{\mu\nu}~~
\sigma_\mu \sigma_\nu^\dagger +\sigma_\nu \sigma_\mu^\dagger = 2\delta_{\mu\nu}
\end{equation}
which implies $W^\dagger W = -\partial_\mu \partial_\mu =
-\partial^2\pmatrix{1&0\cr0&1}$. Thus, one can think about $W$ as
a complex square root of the Laplacian. Similarly, one has
$D^\dagger D=DD^\dagger =-D^2$, with $D^2$ being $-\partial^2$
times a $4\times 4$ unit matrix.

When we deal with gauge theories there are other important
fields~\cite{Luis}. These are the gauge fields, which define a Lie
algebra valued one-form on the torus, denoted by $A\equiv A_\mu
dx_\mu$. We shall take $A_\mu (x)$ to be an anti-hermitian,
traceless, $N\times N$ matrix. The 1-form defines parallel
transport of $N$-component complex fields $\Phi$ by:
\begin{equation}
\Phi (x(1)) = {\cal P} e^{\int_{\cal C} A\cdot dx } \Phi (x(0))
\end{equation}
where $x_\mu (t), t\in[0,1]$ is a curve ${\cal C}$ connecting
$x(0)$ to $x(1)$ and ${\cal P}$ denotes path ordering, the ordered
product of $N\times N$ matrices being implicit in the exponential
symbol. Covariant derivatives, $D_\mu = \partial_\mu -A_\mu$, have
as main property the transformation rule:
\begin{equation}
g^\dagger (x) D_\mu (A) g (x) = D_\mu (A^g )~~~ A^g\equiv A- g^\dagger dg
\end{equation}
where the $g(x)$ are unitary $N\times N$ matrices with unit
determinant. The replacement of $\partial_\mu$ by $D_\mu$ is known
as the principle of minimal substitution and defines $A$-dependent
Weyl and Dirac operators. A major role is played by local gauge
transformations, defined by $\psi\to g\psi,\bar\psi\to\bar\psi
g^\dagger$ and $A\to A^g$ where $\psi$ is viewed as a column and
$\bar\psi$ as a row. The gauge transformations make up an infinite
invariance group and only objects that are invariant under this
group are of physical interest. In particular, $S$ itself must be
gauge invariant and the $\psi$ dependent part of it is of the form
$S_\psi=\int_x \bar\psi W\psi$ with $W$ possibly replaced by
$W^\dagger$ or by $D$.

Formally, $W^{-1}$ is gauge covariant and $\det W$ is gauge
invariant. Both the construction of $W$ and of $D$ meet with some
problems: (1) $W$ may have exact ``zero modes'', reflecting a
nontrivial analytical index. The latter is an integer defined as
$\dim{\rm Ker} W^\dagger (A) - \dim{\rm Ker} W (A) $. It is
possible for this integer to be non-zero because the form $A$ is
required to be smooth only up to gauge transformations. The space
of all $A$'s then splits into a denumerable collection of
disconnected components, uniquely labelled by the index. The
integration over $A$ is split into a sum over components with
associated integrals restricted to each component. (2) $\det W$
cannot always be defined in a gauge invariant way, but
$\det(W^\dagger W)=|\det W|^2$ can. Thus, $\det W$ is to be viewed
as a certain square root of $|\det W|^2$, but, instead of being a
function over the spaces of $A$ it is a line bundle. As a line
bundle it can be also viewed as a line bundle over the space of
gauge orbits of $A$, where a single orbit is the collection of all
elements $A^g$ for a fixed $A$ and all $g$. The latter bundle may
be twisted, and defy attempts to find a smooth gauge invariant
section. When this happens we have an anomaly.

\subsection{Why is there a problem on the lattice ?}

Lattice field theory~\cite{Creutz} tries to construct the desired
functional integral by first replacing space-time by a finite,
uniform, toroidal square lattice and subsequently constructing a
limit in which the lattice spacing, $a$, is taken to zero. Before
the limit is taken functional integration is replaced by ordinary
integration producing well defined quantities. One tries to
preserve as much as possible of the desired symmetry, and, in
particular, there is a symmetry group of lattice gauge
transformations given by $\prod_x SU(N)$, where $x$ denotes now a
discrete lattice site.

The one-form $A$ is replaced by a collection of elementary
parallel transporters, the link matrices $U_\mu(x)$, which are
unitary and effect parallel transport from the site $x$ to the
neighboring site to $x$ in the positive $\mu$ direction. Traversal
in the opposite direction goes with $U_\mu^\dagger (x)$. The
fields $\bar\psi$ and $\psi$ are now defined at lattice sites only.
As a result, $W,W^\dagger$ become finite square matrices. Here are
the main problems faced by this construction: (1) The space of
link variables is connected in an obvious way and therefore the
index of $W$ will vanish always. Indeed, $W$ is just a square
matrix. (2) $\det W$ is always gauge invariant, implying that
anomalies are excluded. In particular, there no longer is any need
to stop the construction at the intermediate step of a line
bundle. These properties show that no matter how we proceed, the
limit where the lattice spacing $a$ goes to zero will not have the
required flexibility.

\subsection{The basic idea of the resolution}

The basic idea of the resolution~\cite{NNplb1} is to reintroduce a
certain amount of indeterminacy by adding to the lattice version a
new infinite dimensional space in which $\psi$ is an infinite
vector, in addition to its other indices. Other fields do not see
this space, and different components of $\psi$ are accordingly
referred to as flavors. Among all fields, only the $\psi$ fields
come in varying flavors. $W$ shall be replaced by a linear
operator that acts nontrivially in the new flavor space in
addition to its previous actions. The infinite dimensional
structure is chosen as simple as possible to provide for,
simultaneously, good mathematical control, the emergence of a
non-zero index and the necessity of introducing an intermediary
construction of $\det W$ as a line bundle~\cite{geom}.

The structure of the lattice $W$ operator is that of a lattice
Dirac type operator. This special lattice Dirac operator, $D$, has
a mass, acting linearly in flavor space. With this mass term, the
structure of our lattice $D$ is:
\begin{equation}
D=\pmatrix{aM^\dagger&aW\cr-aW^\dagger&aM}
\end{equation}
Only $M$ acts nontrivially in flavor space. To obtain a single
Weyl field relevant for the subspace corresponding to small
eigenvalues of $-D^2$, the operator $M$ is required to satisfy:
(1) the index of $M$ is unity (2) the spectrum of $MM^\dagger$ is
bounded from below by a positive number, $\Lambda^2$. $(\Lambda
a)^2$ is of order unity and kept finite and fixed as $a\to 0$. In
practice it is simplest to set the lattice spacing $a$ to unity
and take all other quantities to be dimensionless. Dimensional
analysis can always be used to restore the powers of $a$. In the
continuum, we always work in the units in which $c=\hbar=1$.
Numerical integration routines never know what $a$ is in length
units. The lower bound on $MM^\dagger$ is taken to be of order unity.

The index structure of $M$ ensures that, for eigenvalues of $-D^2$
that are small relative to unity, the relevant space is dominated
by vectors with vanishing upper components. These vectors are
acted on by the $W$ sub-matrix of $D$. Moreover, the main
contribution comes from the zero mode of $M$, so, both the
infinite flavor space and the extra doubling implicit in using a
Dirac operator, become irrelevant for the small eigenvalues of
$-D^2$ and their associated eigenspace.

The standard choice for $M$ stems from a paper by Callan and
Harvey~\cite{CH} which has been ported to the lattice by
Kaplan~\cite{K}. The matrix $M$ is given by a first order
differential (or difference) operator of the form $-\partial_s +
f(s)$, where $s$ is on the real line and represents flavor space.
$f(s)$ is chosen to be the sign function, but could equally well
just have different constant absolute values for $s$ positive and
for $s$ negative.

The construction of the lattice determinant line bundle will not 
be reviewed here and we shall skip ahead directly to the overlap Dirac 
operator. 

\subsection{The overlap Dirac operator}

The continuum Dirac operator combines two independent Weyl
operators. The Weyl components stay decoupled so long as there is
no mass term, and admit independently acting symmetries. Thus,
zero mass Dirac fields have more symmetry than massive ones. In
particular, this implies that radiative corrections to small Dirac
masses must stay proportional to the original mass, to ensure
exact vanishing in the higher symmetry case. A major problem in
particle physics is to understand why all masses are so much
smaller than the energy at which all gauge interactions become of
equal strength and one of the most important examples of a
possible explanation is provided by the mechanism of chiral
symmetry. Until about six years ago it was believed that one could
not keep chiral symmetries on the lattice and therefore lattice
work with small masses required careful tuning of parameters.

Once we have a way to deal with individual Weyl fermions, it must
be possible to combine them pair-wise just as in the continuum and
end up with a lattice Dirac operator that is exactly massless by
symmetry. This operator is called the overlap Dirac operator and
is arrived at by combining the two infinite flavor spaces of each
Weyl fermion into a new single infinite space~\cite{Novl1}.
However, unlike the infinite space associated with each Weyl
fermion, the combined space can be viewed as the limit of a finite
space. This is so because the Dirac operator does not have an
index -- unlike the Weyl operator -- nor does it have an ill
defined determinant. Thus, there is no major problem if the
lattice Dirac operator is approximated by a finite matrix. The two
flavor spaces are combined simply by running the coordinate $s$
first over the values for one Weyl component and next over the
values for the other Weyl component. Since one Weyl component
comes as the hermitian conjugate of the other it is no surprise
that the coordinate $s$ will be run in opposite direction when it
is continued. Thus, one obtains an infinite circle, with a
combined function $f(s)$ which is positive on half of the circle
and negative on the other. The circle can be made finite and then
one has only approximate chiral symmetry~\cite{Napprox}. One can
analyze the limit when the circle goes to infinity and carry out
the needed projection on the small eigenvalue eigenspaces to
restrict one to only the components that would survive in the
continuum limit. The net result is a formula for the lattice
overlap Dirac operator, $D_o$~\cite{Novl1}.

To explain this formula one needs, as a first step, to introduce
the original lattice Dirac operator due to Wilson, $D_W$. That
matrix is the most sparse one possible with the right symmetry
properties, excepting chiral symmetry. It is used as a kernel of
the more elaborate construction needed to produce produce $D_o$.
Any alternative to $D_W$ will produce, by the same construction, a
new $D_o$, possibly enhancing some of its other properties. The
original $D_o$ is still the most popular, because the numerical
advantage of maximal sparseness of $D_W$ has proven hard to beat
by benefits coming from other improvements. Thus, we restrict
ourselves here only to $D_W$.
\begin{eqnarray}
&D_W=m+4-\sum_\mu V_\mu \nonumber\\
&V_\mu=\frac{1-\gamma_\mu}{2} T_\mu +\frac{1+\gamma_\mu}{2}
T_\mu^\dagger\nonumber\\
&\langle x | T_\mu | \Phi^i\rangle = U_\mu (x)^{ij} \langle x |
\Phi^j \rangle\nonumber\\
&U_\mu (x)U_\mu^\dagger(x)=1~~~~\gamma_\mu=\pmatrix{0&
\sigma_\mu\cr\sigma_\mu^\dagger&0}~~~\gamma_5=\gamma_1\gamma_2\gamma_3\gamma_4
\end{eqnarray}
It is easy to see that $V_\mu V_\mu^\dagger = 1$, so $D_W$ is
bounded. $H_W =\gamma_5 D_W$ is hermitian and sparse. The
parameter $m$ must be chosen in the interval $(-2,0)$, and
typically is around $-1$. For gauge fields that are small, the
link matrices are close to unity and a sizable interval around
zero can be shown to contain no eigenvalues of
$H_W$~\cite{Nbounds}. This spectral gap can close for certain
gauge configurations, but these can be excluded by a simple local
condition on the link matrices. When that condition is obeyed, and
otherwise independently on the gauge fields, all eigenvalues of
$H_W^2$ are bigger than some positive number $\mu^2$. This makes
it possible to unambiguously define the sign function of $H_W$,
$\epsilon (H_W)$. As a matrix, $\epsilon$ is no longer sparse,
but, for $\mu^2 >0$, it still is true that entries associated with
lattice sites separated by distances much larger than
$\frac{1}{\mu}$ are exponentially small.

The exclusion of some configurations ruins the simple connectivity
of the space of link variables just as needed to provide for a
lattice definition of the integer $n$, which in the continuum
labels the different connected components of gauge orbit space.
The appropriate definition of $n$ on the lattice is~\cite{top}
\begin{equation}
n=\frac{1}{2} Tr \epsilon (H_W)
\end{equation}
It is obvious that it gives an integer since $H_W$ must have even
dimensions as is evident from the structure of the
$\gamma$-matrices. Moreover, it becomes very clear why
configuration for which $H_W$ could have a zero eigenvalue needed
to be excised. These configurations were first found to need to be
excised when constructing the lattice version of the $\det W$ line
bundle.

The overlap Dirac operator is
\begin{equation}
D_o =\frac{1}{2} (1+\gamma_5 \epsilon (H_W))
\end{equation}
$\gamma_5$ and $\epsilon$ make up a so caller ``Kato pair'' with
elegant algebraic properties~\cite{Kato}.

\subsection{What about the Ginsparg-Wilson relation ?}

In practice, the inverse of $D_o$ is needed more than $D_o$
itself. Denoting $\gamma_5 \epsilon (H_W)=V$, where $V$ is unitary
and obeys ``$\gamma_5$-hermiticity'', $\gamma_5
V\gamma_5=V^\dagger$, we easily prove that
$D_o^{-1}=\frac{2}{1+V}$ obeys
\begin{equation}
\{\gamma_5 , D_o^{-1} -1\} =0
\end{equation}
Here, we introduced the anti-commutator $\{a,b\}\equiv ab+ba$. In
the continuum, the same relation is obeyed by $D^{-1}$ and
reflects chiral symmetry. We see that a slightly altered
propagator will be chirally symmetric. The above equation, 
modifying the continuum
relation $\{\gamma_5 , D^{-1}\}=0$, was first written down by
Ginsparg and Wilson (GW) in 1982~\cite{GW} in a slightly different form. 
By a quirk of history,
their paper became famous only after the discovery of $D_o$.
The main point of the GW paper is that shifting an explicitly chirally symmetric
propagator by a matrix which is almost diagonal in lattice sites and unity
in spinor space does not destroy physical chiral symmetry.

It turns out that the explicitly chirally symmetric propagator, $\frac{1-V}{1+V}$, can be
used as the propagator\ associated with the monomials of the
fields that multiply $e^S$, but in other places where the
propagator appears (loops), one needs to use the more subtly chirally 
symmetric propagator, $D_o^{-1}=\frac{2}{1+V}$. 
This dichotomy is well understood and leads to no
inconsistencies~\cite{Intext}.

Any solution of the GW relation, if combined with $\gamma_5$ hermiticity, 
is of the form $\frac{2}{1+V}$, producing a propagator which anti-commutes 
with $\gamma_5$ of the form $\frac{1-V}{1+V}$. $V$ is a unitary, $\gamma_5$-hermitian,
matrix. Thus the overlap is, essentially, the general
$\gamma_5$-hermitian solution to the GW relation.
The overlap goes beyond the GW paper in providing a generic procedure to produce
explicit acceptable matrices $V$ starting from explicit matrices of the same type
as $H_W$. 

When the GW
relation was first presented, in 1982, the condition of
$\gamma_5$-hermiticity was not mentioned. The solution was not
written in terms of a unitary matrix $V$, and there was no
explicit proposal for the dependence of the solution on the gauge
fields. For these reasons, the paper fell into oblivion, until
1997, when $D_o$ was arrived at by a different route. With the
benefit of hindsight we see now that it was a mistake not to
pursue the GW approach further.

In 1982
neither the mathematical understanding of anomalies - specifically
the need to find a natural $U(1)$ bundle replacing the chiral
determinant - nor the paramount importance of the index of the the
Weyl components were fully appreciated. Only after these
developments became widely understood did it become possible to
approach the problem of lattice chirality from a different angle
and be more successful at solving it. The convergence with the
original GW insight added a lot of credence to the solution and
led to a large number of papers based on the GW relation.

Already in 1982 GW showed that if a solution to their relation
were to be found, the slight violation of anti-commutativity with
$\gamma_5$ that it entailed, indeed was harmless, and even allowed for
the correct reproduction of the continuum triangle diagram, the
key to calculating anomalies. Thus, there was enough evidence in
1982 that should have motivated people to search harder for a
solution, but this did not happen. Rather, the prevailing opinion
was that chirality could not be preserved on the lattice, and
several ``experts'' made careers out of consolidating this view. In
retrospect, something did go wrong in the field's collective
thought process, but parallel developments mentioned earlier
eventually provided an alternative, a ``second chance'' to deal
with the problem correctly. Luckily, despite much opposition, this
second chance was not missed. After the discovery of $D_o$,
fifteen years after the GW paper, a flood of new papers,
developing the GW approach further, appeared. However, nothing
truly new came out from this, because the overlap development
already had produced all the new conceptual results. But, this
renewed activity did provide enough of a social basis in the field
to finally eradicate the misplaced ``wisdom'' of the intervening
years.

\subsection{Basic implementation} Numerically the problem is to
evaluate $\epsilon(H_W)$ on a vector, without storing it, basing
oneself on the sparseness of $H_W$. This can be done because,
possibly after deflation, the spectrum of $H_W$ has a gap around
$0$, the point where the sign function is discontinuous. In
addition, since $H_W$ is bounded we need to approximate the sign
function well only in two disjoint segments, one on the positive
real line and the other its mirror image on the negative side. A
convenient form is the Higham representation, which introduces
$\epsilon_n (x)$ as an approximation to the sign function:
\[
\epsilon_n (x)=\left \{ \begin{array}{ll} \tanh [2n\tanh^{-1}
(x)]&\mbox{for $|x|<1$}\\\tanh [2n\tanh^{-1} (x^{-1})]&\mbox{for $|x|>1$}\\
x&\mbox{for $|x|=1$}
\end{array}
\right. \]
Equivalently,
\begin{equation}
\epsilon_n (x) =\frac {(1+x)^{2n} -(1-x)^{2n}}{(1+x)^{2n}
-(1-x)^{2n}}=\frac{x}{n} \sum_{s=1}^n \frac{1}{x^2\cos^2
\frac{\pi}{2n} \left (s-\frac{1}{2}\right ) + \sin^2
\frac{\pi}{2n} \left (s-\frac{1}{2}\right )}
\end{equation}
\[
\lim_{n\to\infty} \epsilon_n (x) = {\rm sign} (x)
\]

 $\epsilon_n (H_W) \psi$ can be evaluated using a single Conjugate
 Gradient (CG) iteration with multiple shifts for all the pole terms
 labelled by $s$ above~\cite{Implement}.
 The cost in operations is that of a single CG together
 with an overhead that is linear in $n$ and eventually dominates.
 The cost in storage is of $2n$ large vectors. The pole
 representation can be further improved using exact formulae due
 to Zolotarev who, essentially, was able to solve the Remez
 problem analytically for this case. However, for so called
 quenched simulations, where one replaces $\det D_o$ by unity in
 the functional integration, the best is to use a double pass~\cite{doublepass}
 version introduced a few years ago but fully understood only
 recently~\cite{twchiu}. In the double pass version storage and number of
 operations become $n$-independent for large $n$, which, for
 double precision calculations means an $n$ larger than $30$
 or so. Thus, the precise form of the pole approximation becomes
 irrelevant and storage requirements are modest. In
 ``embarrassingly parallel'' simulations this is the method of
 choice because it simultaneously attains maximal numerical
 accuracy and allows maximal exploitation of machine cycles.

 When one goes beyond the $\det D_o =1$ approximation, one needs
 to reconsider methods that employ order $n$ storage. A discussion
 of the relevant issues in this case would take us beyond the
 limits of this presentation; these issues will be covered by
 other speakers who are true experts.

\section{Beyond Overlap/GW ?}

The overlap merged with GW because both ideas exploited a single
real extra coordinate. The starting point of the overlap
construction however seems more general, since it would allow a
mass matrix in infinite flavor space even if the latter were
associated with two or more coordinates. Thus, one asks whether
using two extra coordinates might lead to a structurally new
construction~\cite{ctm}. While this might not be better in practice, it at
least has the potential of producing something different,
unattainable if one just sticks to the well understood GW track.

The function $f(s)$ from the overlap is replaced now by two
functions $f_1 (s_1)$ and $f_2 (s_2)$ and the single differential
operator $\partial_s + f(s)$ by two such operators, $d_\alpha
=\partial_\alpha +f_\alpha (s_\alpha )$. Clearly, $d_1$ and $d_2$
commute. A  mass matrix with the desired properties can be now
constructed as follows:
\begin{equation}
M=\pmatrix{d_1&-id_2^\dagger\cr id_2 & -d_1^\dagger}
\end{equation}
The two dimensional plane spanned by $s_\alpha$ is split into four
quadrants according to the pair of signs of $f_\alpha$ and,
formally, the chiral determinant can be written as the trace of
four Baxter Corner Transfer Matrices,
\begin{equation}
{\rm chiral~det} =Tr [ K^{\rm\bf I} K^{\rm\bf II}K^{\rm\bf
III}K^{\rm\bf IV}]
\end{equation}
While this structure is intriguing, I have made no progress yet on
understanding whether it provides a natural definition of a $U(1)$
bundle with the right properties. If it does, one could go over to
the Dirac case, and an amusing geometrical picture seems to
emerge. It is too early to tell whether this idea will lead
anywhere or not.

\section{Localization and Domain Wall Fermions}

\subsection{What are Domain Wall Fermions ?}

Before the form of $D_o$ was derived we had a circular $s$ space
with $f(s)$ changing sign at the opposite ends of a diameter. One
of the semi-circles can be eliminated by taking $|f(s)|$ to
infinity there, leaving us with a half circle that can be
straightened into a segment with two approximate Weyl fields
localized at its ends. This is known as the domain wall setup, the
walls extending into the physical directions of space-time.
Keeping the length of the segment finite but large one has
approximate chiral symmetry and an operator $D_{DW}$ which acts on
many Dirac fields, exactly one of them having a very small
effective mass, and the rest having masses of order unity.

The chiral symmetry is only approximate because matrix elements of
$\frac{1}{D_{DW}^\dagger D_{DW} }$ connecting entries associated
with opposite ends of the segment, $L$ and $R$, do not vanish
exactly. Using a spectral decomposition of $D_{DW}^\dagger D_{DW}$
we have:
\begin{equation}
\langle L |\frac{1}{D_{DW}^\dagger D_{DW} }|R \rangle = \sum_n
\frac{1}{\Lambda_n} \langle \Psi_n |R\rangle \langle \Psi_n |
L\rangle^*~~~~\langle\Psi_n | \Psi_n \rangle =1
\end{equation}

Weyl states are localized at $L$ and $R$ and should not connect
with each other. So long as the distance between $R$ and $L$ is
infinite and $H_W^2 > \mu^2$ this is exactly proven to be the
case. For a finite distance $S$, the correction goes as $e^{-\mu
S}$. Unfortunately, $\mu$ can be very small numerically and this
would require impractically large values of $S$. Note that the worse 
situation occurs if one has simultaneously a realtively large 
wave-function contribution, $|\langle \Psi_n |R\rangle \langle \Psi_n |
L\rangle |$, and a small $\Lambda_n$. Unfortunately, this worse
case seems to come up in practice.

\subsection{The main practical problem}

As already mentioned, for the purpose of
keeping track of $\det D_o$, one may want to
keep in the simulation the dependence on the coordinate $s$, 
or, what amounts to a logical equivalent, the $n$ fields
corresponding to the pole terms in the sign function truncation.
This is the main reason to invest resources in domain wall 
simulations. In my opinion, if one works in the approximation where $\det D_o =1$  
it does not pay to deal with domain wall fermions because it is
difficult to safely assess the magnitude of chirality violating effects
in different observables. 

The main problem faced by practical domain wall simulations is
that in the range of interest for strong interaction (QCD)
phenomenology $H_W$, the kernel of the overlap, has eigenstates
with very small eigenvalues in absolute value. It turns out that
these states are strongly localized in space-time. However,
because of approximate translational invariance in $s$ they
hybridize into delocalized bands into the extra dimension. As
such, they provide channels by which the Weyl modes at the two
locations $L$ and $R$, where the combined $f(s)$ vanishes, 
communicate with each other, spoiling the chiral symmetry. 
To boot, these states have small $\Lambda_n$. The one
way known to eliminate this phenomenon is to take the separation
between the Weyl modes to infinity. This leads to the overlap
where the problem becomes only of a numerical nature and is
manageable by appropriately deflating $H_W$ to avoid the states
for which the reconstruction of the sign function by iterative
means is too expensive. 

\subsection{The new idea}

The new idea is to exploit the well known fact that one
dimensional random systems typically always localize. The standard
approach uses a homogeneous $s$ coordinate; translations in $s$
would be a symmetry, except for the points at the boundary.
Suppose we randomized to some degree the operators $H_W$ strung
along the $s$-direction, randomly breaking $s$-translations also
locally. This would evade hybridization, making localized states
in the $s$ direction. Now the hope is that the right amount of
disorder would affect only the states made out of the eigenstates
of $H_W$ that are localized in the space-time direction because
there states have small eigenvalues making the basis for the
hybridized states decay slowly in the $s$ direction.

The problem boils down to invent the right kind, and discover the
right amount, of randomness that would achieve the above. A simple
idea is to randomize somewhat the parameter $m$ in $H_W$ as a
function of $s$. A numerical test of this idea, with a small
amount of randomness, has been carried out together with F.
Berruto, T. W. Chiu and R. Narayanan. It turned out that the
amount of randomness we used was too small to have any sizable
effect. The test did show however that if the randomness is very
small nothing is lost, so we have something we can smoothly modify away from.
However the computational resources needed for a more thorough
experiment are beyond our means, so the matter is left unresolved.

\section{Final words}

Much progress has been attained on the problem of lattice
chirality, both conceptually and in practical implementations. The
old wisdom that symmetry is king in field theory has been again
proven. However, there is room for further progress and new ideas
can still lead to large payoffs; keep your eyes open !

\section{Acknowledgments}
Many thanks are due to the organizers of the
``Third International Workshop on QCD and Numerical Analysis''
for an exciting conference. My research is supported in part by
the DOE under grant number DE-FG02-01ER41165.


\begin{thebibliography}{99}
\bibitem{PandS} M. E. Peskin, D. V. Schroeder, An Introduction to
Quantum Field Theory, Addison-Wesley, (1995).
\bibitem{Luis} L. Alvarez-Gaume, Erice School Math. Phys. 0093 (1985).
\bibitem{Creutz} M. Creutz, Quarks, gluons and lattices, Cambridge University Press, (1983).
\bibitem{NNplb1} R. Narayanan, H. Neuberger, Phys. Let. B309, 344 (1993).
\bibitem{geom} H. Neuberger, Phys. Rev. D59, 085006 (1999); H. Neuberger,
Phys. Rev. D63, 014503 (2001); R. Narayanan, H. Neuberger, Nucl. Phys. B443, 305 (1995).
\bibitem{CH} C. G. Callan, J. A. Harvey, Nucl. Phys. B250, 427 (1985).
\bibitem{K} D. B. Kaplan, Phys. Lett. B288, 342 (1992).
\bibitem{Novl1} H. Neuberger, Phys. Lett. B417, 141 (1998).
\bibitem{Napprox} H. Neuberger, Phys. Rev. D57, 5417 (1998).
\bibitem{Nbounds} H. Neuberger, Phys. Rev. D61, 085015 (2000).
\bibitem{top} R. Narayanan, H. Neuberger, Nucl. Phys. B412, 574 (1994); R. Narayanan, H. Neuberger,
Phys. Rev. Lett. 71, 3251 (1993).
\bibitem{Kato} H. Neuberger, Chin. J. Phys. 38, 533 (2000); T. Kato,
Perturbation theory for linear operators, Springer-Verlag, Berlin, (1984).
\bibitem{Novl2} H. Neuberger, Phys. Lett. B427, 353 (1998).
\bibitem{GW} P. H. Ginsparg, K. G. Wilson, Phys. Rev. D25, 2649 (1982).
\bibitem{Intext} H. Neuberger, Nucl. Phys. Proc. Suppl. 73, 697 (1999).
\bibitem{Implement} H. Neuberger, Phys. Rev. Lett. 81, 4060 (1998).
\bibitem{doublepass} H. Neuberger, Int. J. Mod. Phys. C10, 1051 (1999).
\bibitem{twchiu} Ting-Wai Chiu, Tung-Han Hsieh, hep-lat/0306025 (2003), Phys. Rev. E, to appear.
\bibitem{ctm} H. Neuberger, hep-lat/0303009 (2003).

\end{thebibliography}
%


\end{document}